\pdfoutput=1
\documentclass[conference,10pt]{IEEEtran}

\usepackage{dsfont}
\usepackage{graphicx}
\usepackage{epstopdf}
\usepackage{caption}
\usepackage{subcaption}
\usepackage{hyperref}
\usepackage[cmex10]{amsmath}
\usepackage{amsfonts}
\usepackage{cite}
\usepackage{comment}

\newcommand{\iid}{i.i.d.\ }

\newcommand{\cdf}{c.d.f.\ }
\newcommand{\mini}{\ensuremath \wedge}
\newcommand{\maxi}{\ensuremath \vee}
\newcommand{\E}[1]{\ensuremath \mathbb{E}\left[#1\right]}
\newcommand{\ES}[2]{\ensuremath \mathbb{E}_{#2}\left[#1\right]}
\newcommand{\EC}[2]{\ensuremath \mathbb{E}\left[#1 \, | \, #2 \right]}
\newcommand{\ESS}[2]{\ensuremath \mathbb{E}\left[#1 \II{#2} \right]}
\newcommand{\ECSS}[3]{\ensuremath \mathbb{E}_{#2}\left[#1 \, | \, #3\right]}

\newcommand{\I}{\ensuremath{\mathds{1}}}
\newcommand{\II}[1]{\ensuremath{\mathds{1}\!_{\left\{#1\right\}}}}

\newcommand{\RR}{\ensuremath{\mathbb{R}}}

\newcommand{\PP}[1]{\ensuremath{\mathbb{P} \left( #1 \right)}}

\newcommand{\PC}[2]{\ensuremath{\mathbb{P} \left( #1 \, | \, #2 \right)}}

\newcommand{\rec}[1]{\ensuremath \frac{1}{#1} }

\newcommand{\FSeq}[2]{\ensuremath{#1_1, #1_2, \ldots, #1_{#2}}}

\newcommand{\PPr}{\ensuremath \Gamma}      
\newcommand{\MPPr}                      
           {\ensuremath \widetilde{\Gamma}}  
\newcommand{\PPrMM}{\ensuremath \mu}    
\newcommand{\RV}{\ensuremath Z}         
\newcommand{\dens}{\ensuremath g}       
\newcommand{\dist}{\ensuremath G}       
\newcommand{\LF}
           {\ensuremath \mathcal{L}}
\newcommand{\expLF}[1]
           {\ensuremath \ell 
					  \left[ #1 \right]}    

\newcommand{\UnifD}[2]{\ensuremath{\text{Uniform} \left[ #1, #2 \right] }}
\newcommand{\BetaD}[2]{\ensuremath{\text{Beta} \left( #1, #2 \right) }}

\newcommand{\Doc}{\ensuremath d}        
\newcommand{\Csize}{\ensuremath C}      
\newcommand{\Hits}{\ensuremath H}       
\newcommand{\HitsDoc}{\ensuremath \Hits_{\Doc}} 
\newcommand{\MeanHitsDoc}{\ensuremath \overline{\Hits}_{\Doc}} 
\newcommand{\Req}{\ensuremath \Theta}   
\newcommand{\NReqs}{\ensuremath n}      
\newcommand{\DocSpan}{\ensuremath \tau} 
\newcommand{\CatPP}
           {\ensuremath \Gamma}         
\newcommand{\MarkedCatPP}               
    {\ensuremath \widetilde{\Gamma}}    
           
\newcommand{\CatArr}{\ensuremath a}     
\newcommand{\CatArrRate}
           {\ensuremath \gamma}         
\newcommand{\DocReqPP}
           {\ensuremath \mathcal{R}}    
\newcommand{\DocRateFn}
           {\ensuremath \Lambda}        
\newcommand{\DocReqRate}
           {\ensuremath \lambda}        

\newcommand{\FRPP}{\ensuremath \Phi}       

\newcommand{\DDMeanFn}
           {\ensuremath \Xi}            

\newcommand{\DD}{\ensuremath X}         

\newcommand{\iReq}{\ensuremath r}       

\newcommand{\DDFPT}[1]
           {\ensuremath T_{#1}}         
\newcommand{\CharTime}[1]
           {\ensuremath t_{#1}}         

\newcommand{\WinSize}{\ensuremath W}

\newcommand{\SNobj}{\ensuremath N}  
\newcommand{\Est}[1]{\ensuremath \widehat{#1}} 

\newtheorem{thm}{Theorem}

\newtheorem{lem}[thm]{Lemma}
\newtheorem{pro}[thm]{Proposition}

\newtheorem{property}[thm]{Property}

\begin{document}

\title{Catalog Dynamics: Impact of Content Publishing and Perishing
  on the Performance of a LRU Cache}
%
%
%
%
%

\author{
\IEEEauthorblockN{Felipe Olmos\IEEEauthorrefmark{1},
Bruno Kauffmann\IEEEauthorrefmark{2},
Alain Simonian\IEEEauthorrefmark{2} and 
Yannick Carlinet\IEEEauthorrefmark{2}}
\IEEEauthorblockA{\IEEEauthorrefmark{1}Orange Labs and CMAP,
  Email: \textit{luisfelipe.olmosmarchant@orange.com}}
\IEEEauthorblockA{\IEEEauthorrefmark{2}Orange Labs, Email: \textit{firstname.lastname@orange.com}}
}

\maketitle

\begin{abstract}
The Internet heavily relies on Content Distribution Networks and transparent
caches to cope with the ever-increasing traffic demand of users. 
Content, however, is essentially versatile: once published at a given time, its
popularity vanishes over time. All requests for a given document are then
concentrated between the publishing time and an effective perishing time.

In this paper, we propose a new model for the arrival of content requests,
which takes into account the dynamical nature of the content catalog.
Based on two large traffic traces collected on the Orange network, we
use the semi-experimental method and determine invariants of the content
request process. This allows us to define a simple mathematical model for
content requests; by extending the so-called ``Che approximation'', we then
compute the performance of a LRU cache fed with such a request process,
expressed by its hit ratio. We numerically validate the good accuracy of our
model by comparison to trace-based simulation.
\end{abstract}





\section{Introduction}


Driven by video streaming, Internet data traffic is rapidly growing, up to 41\% at the busy hour in 2012 according to a Cisco forecast. Content delivery networks (CDNs) are now a key component of the Internet architecture and play a central role in coping with such a demand. By means of caching and duplicating content near the end-users, CDNs provide an Internet experience with high performance and high availability. Additionally, as the cost of memory decreases faster than that of bandwidth, Internet Service Providers (ISPs) also locally resort to transparent caching to decrease the
load on specific expensive links. 
This favorable bandwidth-memory trade-off has been confirmed by recent research 
\cite{KauffmannITC2013,LessPainMostGain,Westphal_ITC2013}. As
most practical replacement policies have a behaviour similar to
that of Least-Recently-Used (LRU), we here  follow
\cite{LessPainMostGain} in using the LRU replacement policy as a
representative one.  

Video delivery is now the majority of traffic at the busy hour, simultaneously driven by User-Generated Content (UGC) traffic and
by Video-on-Demand (VoD) services. YouTube, the best known of UGC sites today, has indeed emerged as an hyper-giant among the Content Providers, serving up to 25\% of the traffic at the busy hour in ISP networks. On the other side,
Video-on-Demand is growing rapidly, as examplified by the
development of NetFlix. 
Understanding the performance of caches for video
delivery becomes therefore crucial for network provisioning and
operation. It allows simultaneously to decrease the network load,
reduce dimensioning needs and decrease peering costs. 

The versatility of content, however, raises
a new challenge. As new content is continuously published, and (part of) the old
content becomes outdated and non-popular, the popularity of a given document
dynamically evolves with time. The dynamics of the content catalog has significant implication for caching performance. First, even with infinite memory, caches
cannot manage to serve every request: in fact, the first request
for any document will obviously not find the content present in the
cache. Secondly, the request traffic is not
stationary and caches never experience a
steady state; indeed, the set of documents which are currently
stored in the caches slowly evolves with time, as new content
replaces the older one. 

Moreover, as detailed in Section \ref{sec:semiExperiments},
the stationarity periods of the content requests process prove
to be short. Consequently, due to the heavy tail of content popularity
distributions, estimating the popularity 
of content during such a short period leads to significant
variance. Additionally, the cache may not reach its steady
state over short periods, and its performance  will therefore depend on
the recent past. Characterizing the cache performance at the busy
hour is therefore a difficult task due to such inherent
variance. This consequently leads us to express the cache
performance as the long term average hit ratio, which estimates
the average dimensioning gains and fully characterizes the
peering gains.  

In this paper, we provide a first answer to that dynamicity issue. Through basic manipulations, hereafter called \emph{semi-experiments}, on two large traces of YouTube and VoD traffic collected from the Orange
network,  we determine the key invariants of the video request process and propose a model which captures them. This model is amenable to mathematical analysis: Using the
so-called Che approximation \cite{CheApprox}, we express the hit ratio of a
LRU cache fed by such a request process as a function of
basic document statistics. We finally show via simulation that this
approximation accurately matches the empirical hit ratio.

Our key findings are the following:
\textbf{\emph{(i)}} The document arrival process can be well represented by a
  Poisson point process;
\textbf{\emph{(ii)}} The document requests process can be well represented by a
  Poisson-Poisson cluster process;
\textbf{\emph{(iii)}} The hit ratio can be expressed in terms of the distribution of document request intensities and document lifespans only.

The remainder of the paper is organized as follows. Section~\ref{sec:related} presents related work. 
Section~\ref{sec:data} describes the dataset 
and the statistics drawn from traces.  In Section~\ref{sec:semiExperiments}, we apply the semi-experiment methodology and determine the structural invariants 
which are relevant for caching.  Based on these observations, we
then build a model (Section~\ref{sec:theoretical}) for the
request process and estimate the hit ratio for a LRU cache. That
estimate is applied to both YouTube and VoD traces and
successfully compared to the empirical hit ratio in Section~\ref{sec:ME}. 


\section{Related Work}
\label{sec:related}


We describe two areas of related work: content-level traffic
characterization and cache performance analysis.



The popularity distribution of documents has been extensively
discussed since the 90's. It has been shown to exhibit a
light-tailed behavior (typically from a Weibull distribution)
when considering the total number of requests for documents over
a long period, and a heavy-tailed behavior (typically a Zipf
distribution) when analyzing the viewing rate of documents (see
\cite{StretchedExponentialDistribution,kauffmannTrac2012,TonChaYT,Mitra:2011:CWV:1961659.1961662} for recent references).

The temporal pattern of document requests has also been
studied. In \cite{CorrelationPsounis}, authors propose a Markov model with
short term memory for document requests, but the
set of available documents remains fixed. 
The lifespan of documents (defined as the time elapsed
between the first and the last requests) has been studied in
\cite{Cherkasova, YTCaseStudy}. Similarly, articles
\cite{ReplacementPoliciesProxy,Mitra:2011:CWV:1961659.1961662}
show that only a small portion of the documents are active at
any moment, and that popular documents have a significant turnover. Document 
arrivals into the catalog are identified in \cite{StretchedExponentialDistribution}
and their impact on the maximum achievable hit ratio is discussed. The distribution of
requests for the same document over its lifetime is studied in 
\cite{Cherkasova,TonChaYT},
but the results are aggregated over all documents, and therefore do not lead to a 
model for the request process. To the best of our knowledge,  
\cite{2009ZinkCampusYT} is the single prior work aiming at a full 
description of the request process. Due to the short
duration of the studied traces, however, the inclusion of catalog dynamics
is rather simplistic in the proposed model. Finally,  
\cite{KauffmannITC2013} proposes a simple model for the dynamics
of documents which are requested only once, but the set of
documents with several requests remains fixed. 
In terms of traffic characterization, the closest related work to ours is
\cite{cluster}, although the latter focuses on packet-level traffic
characterization. In particular, the authors introduce the so-called
``semi-experimental methodology'' that we use in
section~\ref{sec:semiExperiments}.

As regards cache performance literature, we here only report the
recent literature focusing on the analytical characterization of
the performance of caches applying to LRU policy. An asymptotic
analysis of the LRU miss rate for either Zipf or Weibull requests distribution, with simple closed-form formulas, is
provided in \cite{JelenkovicAAP1999}. Che et al. \cite{CheApprox}
propose another approximation, which is asymptotically exact for
a Zipf popularity \cite{Jelenkovic2008PER} and also accurate for
other types of distributions \cite{FrickerChe}.  

We are aware of a few works which do not assume
\iid request sequences. In particular,
\cite{Jelenkovic2004DependentRequests} studies the performance of
a LRU cache fed by correlated requests, where the instantaneous
request distribution depends on a stationary modulating Markov
process; the asymptotic performance is identical to that under
IRM, showing that such 
a short-term correlation does not fully capture the content dynamics.  \cite{CorrelationPanagakis} also estimates the performance of
a LRU cache when the requests form a Markov chain, but no
closed-form formula is provided. \cite{Fofack} and
\cite{martina2013unified} provide a theoretical analysis of a
network of LRU cache, when requests 
for a given document form an arbitrary renewal
process; correlation among requests can thus be incorporated, but
the catalog of document remains static.

Reference \cite{KauffmannITC2013} is the first published paper to address, though in a limited way, the
dynamics of content catalog. It provides an asymptotic
performance formula when requests for popular documents follow a 
Zipf law under IRM, and an exogenous stream of unique
requests for an infinite set of ``noise'' documents is added. 
Finally, recent developments \cite{LeonardiSNM,LeonardiTemporalLocality}, though not yet refereed, share the core intuitions of our paper; they also model the document
request arrivals using a shot-noise (or Poisson cluster)
process, where the document popularity profile is
parametrized by both its average popularity and its
lifespan. We differ, however, from these papers by the semi-experiment
methodology that we use in section~\ref{sec:semiExperiments} and
which justifies the model we propose. We also validate our results
on two different traces, corresponding to two different traffic
profiles, namely YouTube and Video-on-Demand traffic. Finally, the long
duration of our traces (respectively, 3 months and 3.5 years)
enables us to better emphasize the impact of temporal locality on
the request process.


\section{Dataset}
\label{sec:data}


\subsection{Data Collection}

We have gathered two datasets from two services, which have different traffic
profiles.

The first dataset, hereafter named \textbf{YT}, captures YouTube
traffic of Orange 
customers located in Tunisia. We have access to the logs of a transparent
caching system set up in order to offload the country's international
connection. This system is a commercial product from a large company
specialized in the design and management of CDNs. In the observation period of
January--March 2013, we collected  around $420\,000\,000$ requests from
about $40\,000$ IP addresses to $6\,300\,000$ chunks.  For each
chunk request in this trace, the logs contain the user (anonymous) IP
address, a video identifier, the timestamp of the end of session, the number of
transmitted bytes, the duration of the HTTP connection and the beginning and
ending position of the specific \emph{chunk} requested, the latter information
being available for $96\%$ of the data. 

The second dataset, hereafter called \textbf{VoD}, comes from the Orange
Video-on-Demand service in France. This service proposes to Orange customers
both free catch-up TV programs and pay-per-view films and series episodes.
Probes deployed at the access of the service platforms recorded video requests
from June 2008 to November 2011. The data amounts to more than $3\,400\,000$ 
requests from $60\,000$ users to $120\,000$ videos.  The records in this
trace consist of the request timestamp, an internal client (anonymous)
identifier and a video identifier.


\subsection{Processing}
For simplicity and mathematical tractability, we focus our analysis at the
document level rather than chunk level.  

Since the YT trace consist of chunk requests, we consolidate them to identify
the user video sessions. To this aim, we first identify the requests from a
single user to a single object. Then, when the chunking information is
available, we simply concatenate the requests corresponding to a chunk chain.  
For the requests without chunk identification, we aggregate all the requests
made by the same user for the video, and with inter-arrival time smaller than 8
minutes.  This threshold corresponds to the $95\%$ percentile of the length
session distribution of requests with chunk data.\footnote{The reason to select
a percentile instead of the maximum is that there are 185 chunk chains with
extreme duration (in the order of days or even months), rendering the maximum a
meaningless  aggregation criterion.}
The result of this procedure is our working YT dataset consisting of
more than $46\,000\,000$ requests to around $6\,300\,000$ unique documents.

In the case of the VoD trace there was no need of the above consolidation
procedure. Nonetheless, the trace contained requests to movie trailers or to
full content but with short duration. We consider these content ``surfing''
requests not relevant in terms of caching performance and thus we discarded
them from the VoD dataset. The working VoD dataset contains around $1\,800\,000$
requests to more than $87\,000$ different objects.
\subsection{Distribution of the Number of Requests}

\begin{figure}[b]
\centering
\input{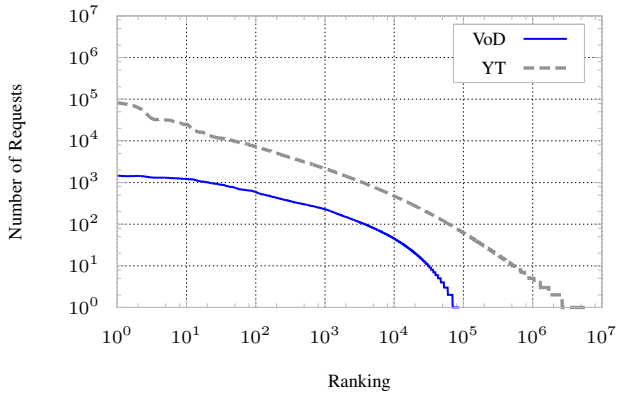}
\caption{\textit{Number of requests as a function of their rank}.}
\label{fig:n_req_fit}
\end{figure}

The logarithmic rank-frequency chart in Figure~\ref{fig:n_req_fit} shows two
different popularity behavior for the traces.  As expected, in the ``short'' YT
trace, the ten thousand most popular documents follow a Zipf distribution with
exponent 0.61, while the tail has an exponent of 1.03.  As for the VoD trace,
the popularity does not follow a power law, but is best fitted by a Weibull
(also called Stretched-Exponential) distribution.

\subsection{Distribution of Lifespan and Request Rate} 
\label{sec:est_quant}

\begin{figure}[t]
\centering
\input{figs/tau_fit.tex} 
\caption{\it Kernel estimate for the lifespan $\DocSpan$.}
\label{fig:tau_pdf}
\end{figure}

\begin{figure*}[t]
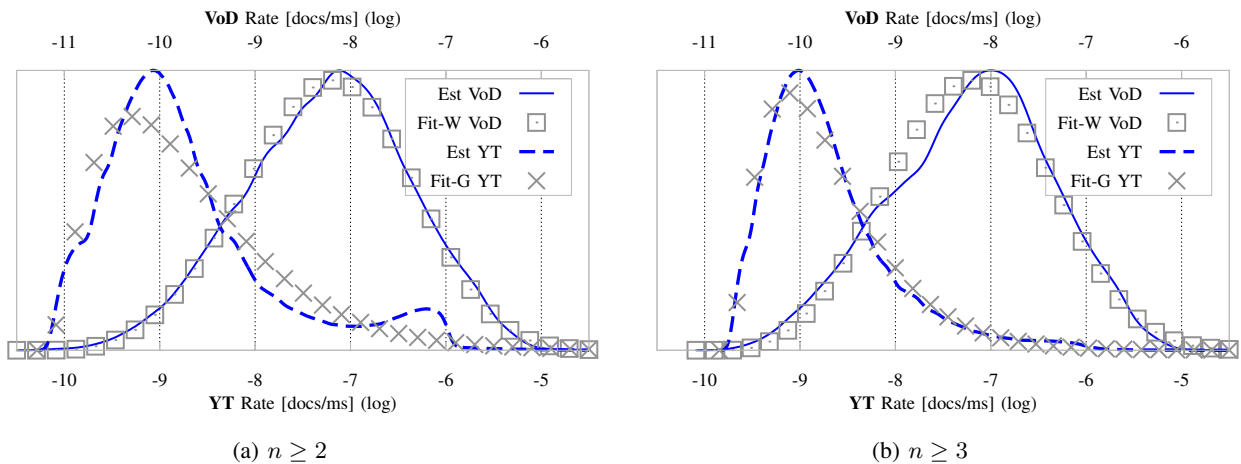

\begin{subfigure}{0.45\textwidth}
\centering
\input{figs/lambda_fit.tex} 
\caption{\it $\NReqs \geq 2$}
\label{fig:lambda_pdf_ge2}
\end{subfigure}%
\quad%
\begin{subfigure}{0.45\textwidth}
\centering
\input{figs/lambda_fit_ge3.tex} 
\caption{\it $\NReqs \geq 3$}
\label{fig:lambda_pdf_ge3}
\end{subfigure}
\caption{\it Kernel estimate for the intensity $\DocReqRate$ and their parametric fittings.}
\label{fig:lambda_pdf}
\end{figure*}
We here provide a finer characterization that considers not only the number of
requests to a given document, but also the period of time during which the
document is active.  For a given document, let $\DocSpan$ denote
its lifespan, that is, the period where the users can 
address requests to it; let then $\DocReqRate$ be the corresponding average
request rate to that document. We estimate these two quantities that form the
basis of our analysis. 
106366702000
Specifically, consider a given document with $\NReqs \geq 2$ requests, and let
$\Req_{\text{I}}$ and $\Req_{\text{F}}$ be the Initial and
Final request times to that document in the observation window. We
then estimate the catalog lifespan $\DocSpan$ by the unbiased estimator
\begin{equation}
\widehat{\DocSpan} = (\Req_{\text{F}} - \Req_{\text{I}}) 
\times  
\frac{\NReqs + 1}{\NReqs - 1}
\label{EstimTau}
\end{equation}
(in fact, assuming that the $\NReqs$ request times $\Theta_1$, ..., $\Theta_n$
are uniformly distributed on the interval with length $\tau$, we easily
calculate $\E{\Theta_F - \Theta_I} = \frac{\NReqs-1}{\NReqs+1} \times
\DocSpan$). 

Regarding intensity $\DocReqRate$, 
our sample is biased by the fact that we collect only documents
with at least one request.
To take this bias into account, we assume that $\NReqs$ is a
Poisson random variable with mean $\DocReqRate \DocSpan$, given $\NReqs
\geq 1$. We thus estimate the request rate $\DocReqRate$ by 
\begin{equation}
\widehat{\DocReqRate} = \NReqs'/\widehat{\DocSpan} 
\label{EstimLambda}
\end{equation} 
where $\NReqs'$ verifies equation $\NReqs'/(1-e^{-\NReqs'}) =
\NReqs$ (the latter is easily shown to have a unique positive solution $n'$;
note in practice that we can take $\NReqs' \approx \NReqs$ for $\NReqs$ greater
than 10). Both estimators $\widehat{\DocSpan}$ and $\widehat{\DocReqRate}$ are
valid only for documents for which we have at least two requests.
Consequently, in the remaining of this section, we will make the analysis over
the set of documents that have at least two requests. 

Figure~\ref{fig:tau_pdf} shows a
kernel density approximation of $\widehat{\DocSpan}$ for each
dataset. Note that the formula of $\widehat{\DocSpan}$ allows 
a positive density for values larger thant the observation
window, especially for documents with a small number of requests.
Also, in the YT data,  we observe a probability mass accumulation
effect near the 
mark of three months, which is precisely the size of the observation window.
This is a truncation effect and it is a sign that the lifespan of a video may
be far longer than our current observation window in this
dataset. 
As regards the VoD data, most documents have a lifespan shorter
than one month. This corresponds to the numerous catch-up TV programs. The
remaining documents have a different distribution, with lifespans
varying on the range of a few weeks to the observation period (3.5
years). Due to the large observation period, the
truncation effect is not visible.

\begin{figure*}[t]
\begin{subfigure}{0.45\textwidth}
\centering
\input{figs/otu_joint_fit.tex}
\caption{\it YT trace.}
\label{fig:otu_joint_fit}
\end{subfigure}%
\quad%
\begin{subfigure}{0.45\textwidth}
\centering
\input{figs/vod_joint_fit.tex}
\caption{\it VoD trace.}
\label{fig:vod_joint_fit}
\end{subfigure}
\caption{\it Joint density kernel estimate for the lifespan/log-intensity
vector $(\DocSpan, \log \DocReqRate)$.}
\label{fig:joint_fit}
\end{figure*}

In the case of $\DocReqRate$, even though its estimator $\widehat{\DocReqRate}$
is biased, we were able to analyze it more throughly than estimator
$\widehat{\DocSpan}$. In fact, with help of a maximum likelihood method, we
have found that a (shifted) Gamma model fit the distribution of $\log
\widehat{\DocReqRate}$ in the YT case and a Weibull model in the VoD case.

In the YT case we observe in Figure~\ref{fig:lambda_pdf} that the fit on the
subsample $\NReqs \geq 3$ is considerably better than on $\NReqs \geq 2$. This
is due an accumulation of mass at the right extreme of the distribution due to
the fact that the estimation of $\DocSpan$ when $\NReqs = 2$ has a lot of
variability and the latter subsample has more than half of the data. Indeed, we
see this effect disappear on the subsample $\NReqs = 3$.

These fittings suggests that, in our traces, the random variable $\DocReqRate$
has a heavy tailed distribution
\footnote{
	We here use the broader definition of ``heavy tail'' random variable $X$ in
	the sense that $\lim_{x \uparrow +\infty} e^{\alpha x} \PP{X > x} = +\infty $
	for every positive $\alpha$. This includes, in particular, power law tailed
	distributions.
}, 
which is consistent with the fact that it is a popularity measure as well. 

Finally the joint distribution of the pair $(\log \widehat{\DocReqRate},
\widehat{\DocSpan})$ is shown in Figure~\ref{fig:joint_fit}, with a focus on
small values of $\widehat{\DocSpan}$ for the VoD data.  In both cases,  we
conclude from the empirical densities that $\DocSpan$ and $\DocReqRate$ are not
independent random variables, and the joint distribution is not easy to fit. We
therefore will use the empirical joint distribution in the following.  Finally,
the presence of managed catch-up TV documents in the VoD data is visible; the
marginal $\widehat{\DocSpan}$ shows density peaks at values of 1, 2 and 4
weeks, corresponding to the duration for which broadcasts remain available.

\section{Semi-Experiments}
\label{sec:semiExperiments}


In this section, we address the identification of  the structural properties of the request process
that are relevant to LRU caching, namely:
\begin{enumerate}
	\item[(i)] Overall correlation between requests.
	\item[(ii)] Correlation in the catalog publications.
	\item[(iii)] Correlation between the requests of a document.
\end{enumerate}
Additionally, in the case of the first property, we look for the timescale
where it starts to influence the performance of LRU.

To this aim, we use the semi-experimental methodology \cite{cluster}. Each
semi-experiment is based on two procedures: The first one is to randomly rearrange
the original request sequence in a way that destroys a specific correlation
structure; the second one is to use this new trace as an input for a simulation of a
LRU cache and compute the corresponding hit ratio curve. We then look at the
discrepancies from the hit ratio curve of the original trace; if they differ
significantly we infer that the broken structure is relevant to LRU caching.
In the following, we explain in detail each semi-experiment and its findings.

\begin{figure}[t]
\centering
\includegraphics[scale=0.33]{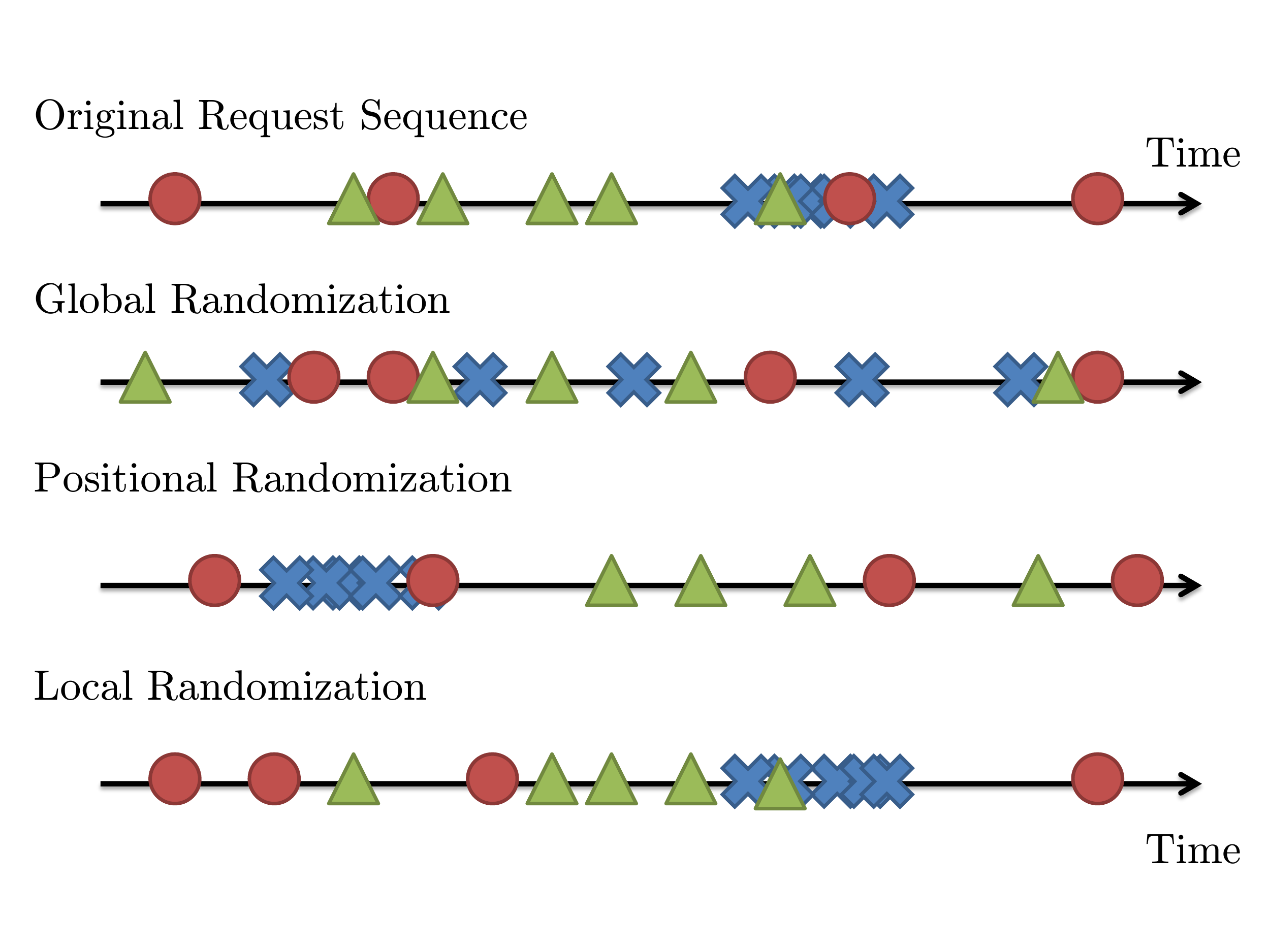}
\caption{\it A schematic view of all three randomizations. 
For global
randomization, all request times are just shuffled. For positional randomization, we
shift the whole request sequence to a random location, preserving the order of
inter-arrival times. For local randomization, we fix the first and last request and
shuffle the times in the middle.
}
\label{fig:randomizations}
\end{figure}

\subsubsection*{Overall Correlation Between Requests}
In this semi-experiment, we completely break the correlation structure of the
request sequence by placing each request at an i.i.d. uniform
time in the interval $\left[0;\WinSize\right]$, where $\WinSize$
is the size of the observation window. 
Any request sequence
shuffled in this manner leads to a IRM sequence, since the process destroys any
dependence structure.
We call this procedure \emph{global
  randomization} and show an example in Figure~\ref{fig:randomizations}.

\begin{figure}[t]
	\begin{subfigure}{1.0\linewidth}
	\input{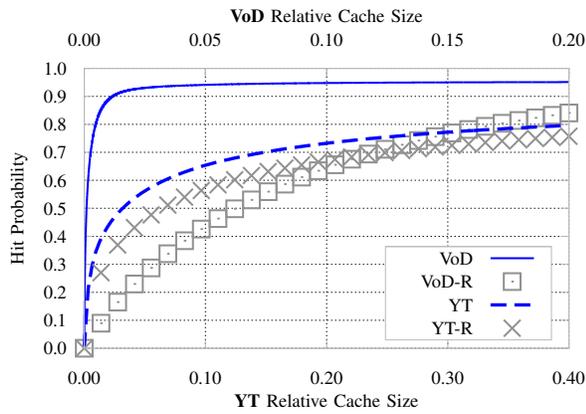}
	\caption{\it Global Randomization}
	\label{fig:orig_vs_glo}
	\end{subfigure}
	
	\begin{subfigure}[b]{1.0\linewidth}
	\input{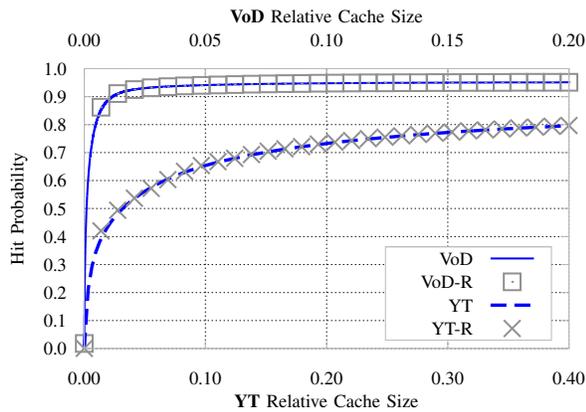}
	\caption{\it Positional Randomization}
	\label{fig:orig_vs_pos}
	\end{subfigure}
	
	\begin{subfigure}[b]{1.0\linewidth}
	\input{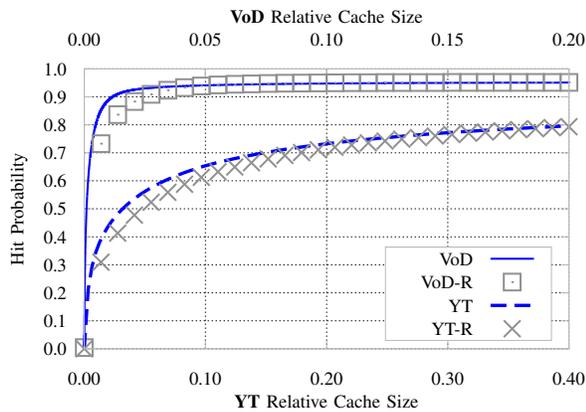}
	\caption{\it Local Randomization}
	\label{fig:orig_vs_loc}
	\end{subfigure}
\caption{\it Comparison of the hit ratio of the original request sequence
versus the results of each randomization.}
\end{figure}

In Figure~\ref{fig:orig_vs_glo}, we compare the resulting hit ratio to that
obtained with the original trace and observe that the hit ratio of the latter
is lower for any cache size in both datasets, but notoriously in the VoD case.
To be more precise, we compute the \emph{mean absolute relative error} or MARE%
\footnote{
The MARE between a model sequence $(y_i)_{1 \leq i \leq N}$ and empirical data
$(x_i)_{1 \leq i \leq N}$ is defined as $\frac{1}{N}\sum_{i=1}^N \frac{| x_i -
y_i |}{|x_i|}$.
} between hit ratio curves of the original and randomized sequence.  In the YT
case, the MARE has a value of 5.0\%; this value might seem low, but it comes
mostly from the left of the curve.  Since the left part of the curve is where
practical cache sizes lie, this discrepancy, however low, is still important.
As for the VoD trace, the MARE amounts to 17.3\% which confirms the huge
difference observed above in Figure~\ref{fig:orig_vs_glo}. 
We thus conclude that the correlation between requests is a meaningful factor
for the performance of LRU caching and that the IRM assumption leads to a
underestimation of the hit ratio, which can be very significant. 

\subsubsection*{Correlation in Catalog Publications}
We now examine how sensitive is our data with respect to the publication of new
documents to the catalog. To this aim, we perform a \emph{positional
randomization}, which breaks the correlation structure between the first
requests of documents, which we use as an estimate of the publication time. 
The procedure consists, for a given document, in leaving the inter-arrival
times of its request sequence unchanged and jointly shift all of them by a
random quantity, as shown in
Figure~\ref{fig:randomizations}. More precisely, let 
$\FSeq{\Req}{k}$ the request times for a document; first, we draw a uniform
random number $U$ from the interval~$[0, \WinSize - (\Req_k - \Req_1)]$, then
we define the new request sequence $\FSeq{\Req^*}{k}$ by $\Req^*_i = U + \Req_i
- \Req_1$ for $1 \leq i \leq k$.

In both traces, the resulting hit ratio shows no difference from the original,
as observed in Figure~\ref{fig:orig_vs_pos}. The MAREs in this semi-experiment
are merely 0.3\% in the YT case and 0.1\% in the VoD case. We therefore
conclude that document arrivals have no correlation structure with
significant impact on caching.

\subsubsection*{Correlation between Requests of a Document}
In this semi-experiment, we aim to break the request dependence structure for each
document. To achieve this, we perform a \emph{local randomization}: For a given
document, we keep its first and the last request times fixed and
only shuffle the 
ones in between at \iid times following a
$\UnifD{\Req_1}{\Req_k}$-distribution. Note that this procedure preserves the
lifespan and intensity statistics discussed in Section~\ref{sec:est_quant}, but
breaks any other correlation structure inherent to the request process of the
document.

Figure \ref{fig:orig_vs_loc} shows that, although the resulting hit ratio is
slightly below (resp. over) the original for small (resp. large) cache sizes,
the MARE is just 1.6\% in the YT trace and 0.7\% in the VoD trace. 
We thus conclude that the correlation among requests of a given document has
little impact on the LRU performance and we can safely neglect it
for modeling purpose.

\subsubsection*{Relation between Correlations and Timescales}

\begin{figure}[t]
	\begin{subfigure}{1.0\linewidth}
	\input{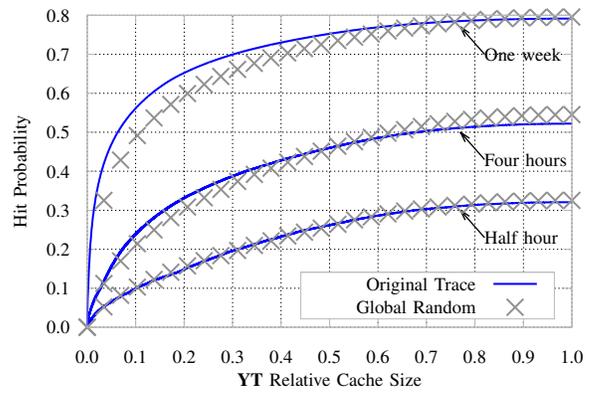}
	\caption{\it YT Trace}
	\end{subfigure}
\vspace*{.2cm}	
	\begin{subfigure}[b]{1.0\linewidth}
	\input{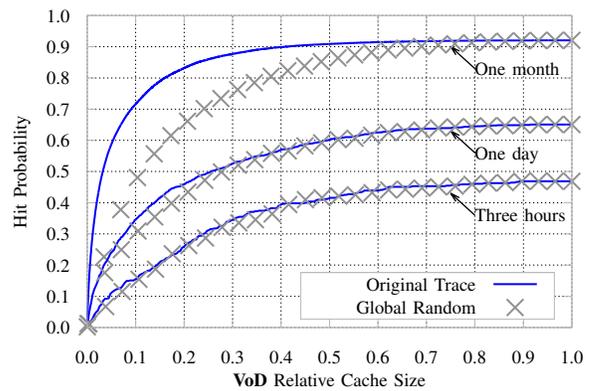}
	\caption{\it VoD Trace}
	\end{subfigure}
\caption{\it Comparison between the hit ratio of the original trace and
the global randomization at different scales.
\label{fig:timescale}
}
\end{figure}

We now determine at which timescale the correlation between requests has an
impact in the LRU performance. With this in mind, we design a slightly
different semi-experiment where we first extract sub-traces of different
timescales, choosing high load periods.
Then we apply the global randomization semi-experiment to each of these
shorter traces. For each dataset, we distinguish three timescales and the
results for each one are shown in Figure~\ref{fig:timescale}; other timescales
lead to results that are just intermediate to the three presented here.

Near the first timescale (one week for YT and one month for VoD) and beyond, all
timescales have a request correlation structure that approaches that
observed in the full trace, and thus its hit ratio differs significantly from
that of the global randomization. Indeed, already at this
timescale, the MAREs
are of 5.3\% and 11.6\% in the YT and VoD datasets, respectively; around the
second timescale (four hours for YT and one day for VoD), we observe a decrease in
the discrepancies as the MAREs  are 5\% and 2.3\% in the YT and VoD
case, respectively. Though we see that the correlation structure does not
influence strongly the hit ratio, we remark again that the underestimation happens in
the left side of the curves which corresponds to practical cache sizes. Finally for
traces around the last timescale (half hour for YT and three hours for VoD),
the MARE are 1.4\% and 2.3\% for YT and VoD respectively and 
we thus 
conclude that there are no significant structures between requests at this
timescale.

\subsubsection*{Insights Gained}
The latter results of the semi-experiments lead us to conclude that: 
\begin{enumerate}
\item[I1:] The correlation structure of the whole request process is not
	negligible, in terms of the hit ratio, at large timescales.  We also infer
	that most of the correlation comes from the fact that all requests for the
	same document are grouped within its lifespan. 

\item[I2:] The document publications exhibit a correlation structure that does
	not have a significant impact on the hit ratio. In particular, we deduce that
	document arrivals to the catalog can be modeled by a Poisson process without
	losing accuracy on the estimation of the hit ratio curve.

\item[I3:] For a given document, the request process within its lifespan
	exhibits some structure, but with little impact of the
        hit ratio. Thus, for a given document, we can approximate
        the requests sequence by a Poisson 
	process defined on the lifespan of the document while
        still preserving the hit ratio. 
\end{enumerate}


\section{Mathematical Analysis}
\label{sec:theoretical}


In this section, we use the previous insights to build a mathematical
model for the whole request process  and detail the estimation of
the corresponding hit rate in a LRU cache 
(throughout, the caching granularity is that corresponding to a document). The
reader can find the proofs of all propositions in the appendix.

\subsection{Catalog Arrival and Request Processes}
\label{CARP}

We build our model for the document request process by following a top-down approach:

- on the top level, we consider the ground process $\CatPP$, hereafter called
\textbf{catalog arrival process}; this point process dictates the consecutive
arrivals of documents to the catalog. In our model, $\Gamma$ is assumed to be
a homogeneous Poisson process with constant intensity
$\CatArrRate$, according to Insight I2.

- let then $\Doc$ be the index of a document generated by process $\CatPP$,
whose arrival time to the catalog is denoted by $\CatArr_\Doc$. Document $\Doc$
then generates a \textbf{document request process} $\DocReqPP_\Doc$
determined by two random variables $\DocRateFn_\Doc$ and $\DocSpan_\Doc$.
Specifically, given $\DocRateFn_\Doc$ and $\DocSpan_\Doc$, we assume
the document request process to be Poisson with intensity function
$\DocRateFn_\Doc$ on interval $[\CatArr_\Doc, \CatArr_\Doc +
\DocSpan_\Doc]$ (cf. Insight I3); the duration $\DocSpan_\Doc$ is the lifespan of document $\Doc$,
intensity function $\DocRateFn_\Doc$ being zero outside interval
$[\CatArr_\Doc, \CatArr_\Doc + \DocSpan_\Doc]$. In the following, we assume that
\begin{equation}
\overline{\NReqs}_\Doc = \int_{\CatArr_\Doc}^{+\infty} \DocRateFn_\Doc(u) \mathrm{d}u
\label{ndbar}
\end{equation}
is almost surely finite;

- finally, the superposition of all processes $\DocReqPP_\Doc$ for all $\Doc$
generates the \textbf{total request process} $\DocReqPP = \Sigma_\Doc
\DocReqPP_\Doc$ that contains the requests to all documents. In the
following, we will also denote by $\DocReqPP'_\Doc = \DocReqPP \setminus
\DocReqPP_\Doc$ the request process resulting from the removal from total
process $\DocReqPP$ of points pertaining to request process
$\DocReqPP_\Doc$ associated with given document $\Doc$.

\begin{figure}
\centering
\includegraphics[scale=0.3]{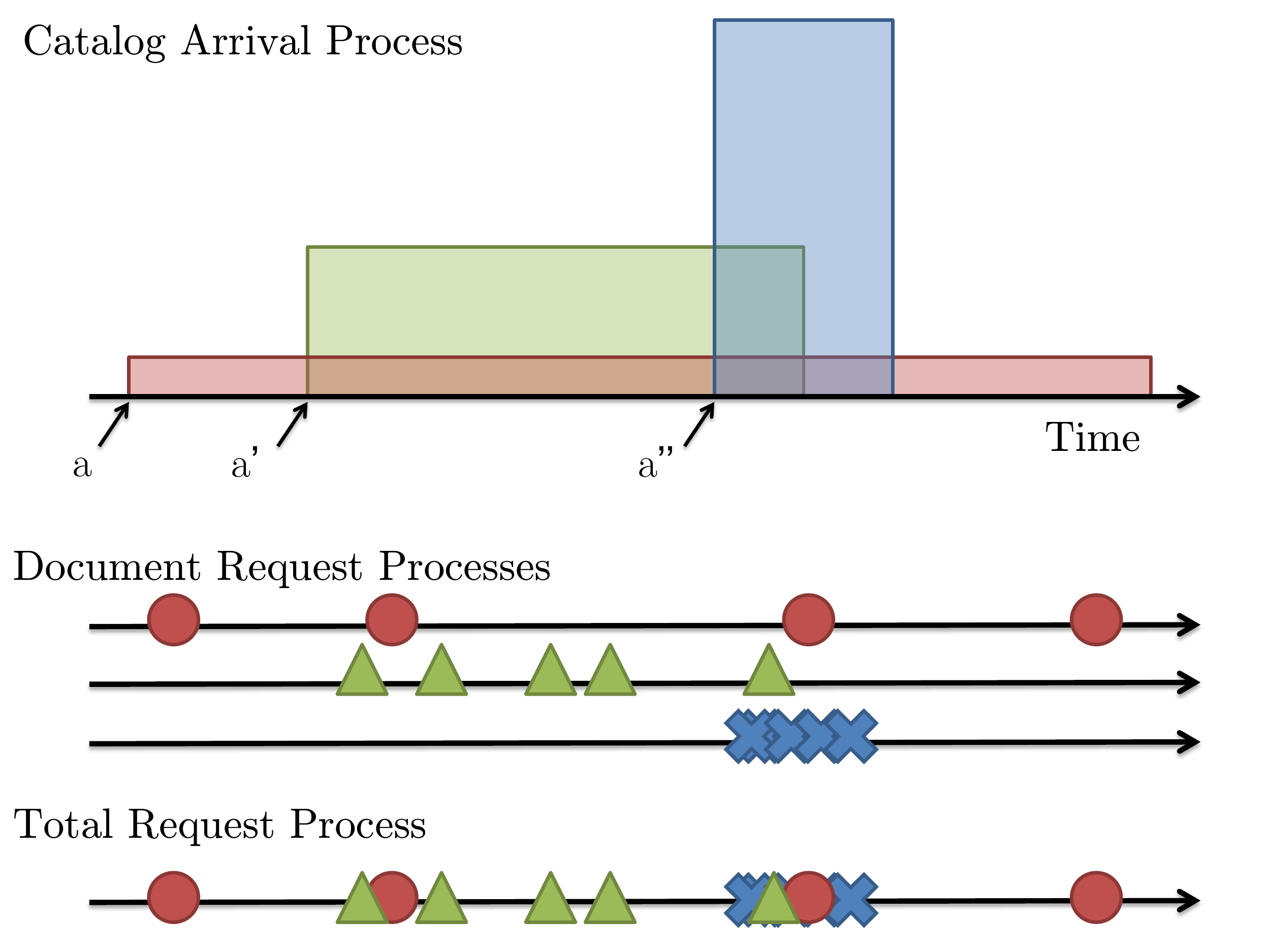}
\caption{\it A sample of the document arrival and request
processes.
\textbf{Top:} The boxes represent the lifespan and popularity by their width
and height, respectively (e.g., the document arriving at $a$ is less popular
than that arriving at $a''$ but it has a longer lifespan).
\textbf{Bottom:} A sample of the document request processes (color coded for
each object). Their superposition generates the total request process.}
\label{fig:request_process}
\end{figure}

We can regard the point process $\DocReqPP$ either as a doubly-stochastic
Poisson process (such processes are also called \textit{Cox} in
the literature) 
which is a Poisson Process with random intensity, in our case the shot-noise
process generated by the popularity functions $\DocRateFn_\Doc$. Additionally, we can regard
$\DocReqPP$ as a \textit{cluster} point process. Figure
\ref{fig:request_process} gives a schematic view of all components of our
request model. 

\subsection{General Hit Ratio Estimation}
\label{TCA}

Given the dynamical request model presented in Section \ref{CARP}, we now
discuss the adaptation of the so-called ``Che approximation'' \cite{CheApprox,
FrickerChe} to calculate the hit ratio for requests addressed to a cache ruled
by the LRU policy. 

Assume that a given document $\Doc$ arrives to the catalog at
time $\CatArr_\Doc$. As the request process to document $\Doc$ is
a Poisson process with intensity function $\DocRateFn_\Doc$, the
sequence $\FSeq{\Req}{\NReqs_\Doc}$ of request times to $\Doc$
has $\NReqs_\Doc$ elements, where $\NReqs_\Doc$ follows a Poisson
distribution with parameter $\overline{\NReqs}_\Doc$ introduced
in (\ref{ndbar}). The expected number of hits to the given document $\Doc$ 
then reads
\begin{align}
\label{eq:EH_begin}
& \MeanHitsDoc 
= 
\ES{\HitsDoc}{\Doc}
=
\ES{\sum_{\iReq = 2}^{\NReqs_\Doc} 
\II{\text{Request at } \Req_\iReq \text { is a hit}}}{\Doc} 
\\
& = \sum_{k = 2}^{+\infty}
\left(
	\sum_{\iReq = 2}^{k} 
	\ECSS{\II{\text{Request at } \Req_\iReq \text { is a hit}}}
	   {\Doc}{\NReqs_\Doc = k} 
\right)
 \frac{e^{-\overline{\NReqs}_\Doc} \overline{\NReqs}_\Doc^k}{k!}
\nonumber
\end{align}
where $\mathbb{E}_\Doc$ denotes the expectation, given $\DocRateFn_\Doc$
and $\DocSpan_\Doc$.
To proceed further with the calculation of $\MeanHitsDoc$, we need to
incorporate the caching management policy. Specifically, we consider a cache of
size $\Csize$ ruled under the LRU policy; the request at time $\Req_\iReq$ will
then be a hit if and only if $\Csize$ different documents have been requested
since the request arrival at time $\Req_{\iReq - 1}$. To formalize this
condition, let
$\DD^s_t$ denote the number of \textit{different} documents requested in time interval $[s, t]$, that is, $\DD^s_t = \# \left\{ \text{Different documents requested in } [s,t] 
\right\}$ 
for $t > s$. From the stationarity of the ground process $\CatPP$, 
we first deduce the following.
\begin{pro}
\label{pro:DD_laws}
For any $s > 0$, processes $(\DD^s_t)_{t \geq s}$ and $(\DD^0_{t - s})_{t \geq 0}$ have identical distributions. Furthermore, $\DD^0$ is a Poisson process with associated mean function $\E{X^0_t} = \DDMeanFn(t)$ given by
\[
\DDMeanFn(t) = \CatArrRate \,
\int_{-\infty}^t
\E{ 1 - \exp{\left\{ - \int_0^t \DocRateFn_\Doc(v)\, dv  \right\}} }
\mathrm{d}\CatArr
\]
for all $t \geq 0$ (in the latter integral, variable $a$ stands for $a_d$ for brevity).
\end{pro}
The latter formula for $\E{X^0_t} = \DDMeanFn(t)$ can be easily interpreted by
noting  that the mean number of different documents arriving in interval
$[a,a+\mathrm{d}a[$ being  $\gamma \mathrm{d}a$ (for any $-\infty < a = a_d <
t$), each of those documents is requested in interval $[0,t]$ with probability
$1 - \exp ( - \int_{[0,t]} \DocRateFn_\Doc(v)\mathrm{d}v )$. 

Now, an immediate consequence of Proposition \ref{pro:DD_laws} is that the first passage
time $\DDFPT{\Csize}^s = \inf \left\{ t \geq s : \DD_t^s = \Csize \right\}$ 
of process $(X^s_t)_{t \geq s}$ to level $\Csize$ has the same distribution than  $\DDFPT{\Csize} + s$, where $\DDFPT{\Csize} = \DDFPT{\Csize}^0$. We can now proceed with the calculation of $\MeanHitsDoc$ expressed in (\ref{eq:EH_begin}). From the previous discussion, a hit event at time $\Req_\iReq$ can be equivalently written as 
$$
\left\{\text{Request at } \Req_\iReq \text{ is a hit} \right\} 
= 
 \left\{ \Req_\iReq - \Req_{\iReq - 1} 
  < \DDFPT{\Csize}^{\Req_{\iReq-1}}  - \Req_{\iReq - 1} \right\}
$$
in terms of $\DDFPT{\Csize}$. Recall that, given the arrival of document $\Doc$ at time $\CatArr_\Doc$, the remaining process $\DocReqPP'_\Doc = \DocReqPP \setminus \DocReqPP_\Doc$ 
has the same distribution than $\DocReqPP$. 
It follows that the distribution of $T_C$ for process $\DocReqPP$ is
identical to that associated with remaining process $\DocReqPP'_\Doc$. As  $\DDFPT{\Csize}^{\Req_{\iReq - 1}} =
\DDFPT{\Csize} + \Req_{\iReq - 1}$ in distribution, we can eventually write (\ref{eq:EH_begin}) as
\begin{equation}
\label{eq:EH_unfolded}
\MeanHitsDoc = 
\sum_{k = 2}^{+\infty}
\left(
	\sum_{\iReq = 2}^{k} 
	\ECSS{\II{\Req_\iReq - \Req_{\iReq - 1} < T_C}}{\Doc}{\NReqs_\Doc = k}
\right)
 \frac{e^{-\overline{\NReqs}_\Doc}\overline{\NReqs}_\Doc^k}{k!}.
\end{equation}

The distribution of $\DDFPT{\Csize}$ intervening in (\ref{eq:EH_unfolded}) is usually unknown or hard to calculate. To overcome 
this difficulty, we now invoke the so-called \textbf{Che approximation}: we
assume that the distribution of $\DDFPT{\Csize}$ is very concentrated so that
it can be approximated by a constant $\CharTime{\Csize}$, hereafter  called the
\emph{characteristic time}. The calculation of that characteristic time then
proceeds as follows; using Proposition \ref{pro:DD_laws}.B together with the
approximation
$\DDFPT{\Csize} \approx \CharTime{\Csize}$, we can write $\Csize = \E{\DD_{\DDFPT{\Csize}}} 
\approx 
\E{\DD_{\CharTime{\Csize}}} = \DDMeanFn(\CharTime{\Csize})$. 
We therefore define the characteristic time $\CharTime{\Csize}$ 
by the inverse relation
\begin{equation}
\CharTime{\Csize} = \DDMeanFn^{-1}(C);
\label{tC}
\end{equation}
replacing $\DDFPT{\Csize}$ by $\CharTime{\Csize}$ in (\ref{eq:EH_unfolded}), we then  obtain the approximation 
\begin{equation}
\label{eq:EH_unfoldedBIS}
\MeanHitsDoc \approx 
\sum_{k = 2}^{+\infty}
\left(
	\sum_{\iReq = 2}^{k} 
	\ECSS{\II{\Req_\iReq - \Req_{\iReq - 1} < t_C}}{\Doc}{\NReqs_\Doc = k}
\right)
 \frac{e^{-\overline{\NReqs}_\Doc} \overline{\NReqs}_\Doc^k}{k!}
\end{equation}
for the expected number of hits.

\subsection{Application to the Box Model}
The general expressions for the average $\DDMeanFn(t) =
\E{X_t^0}$ and the expected number of hits derived in Section
\ref{TCA} are now applied to the specific \textit{Box Model} in
order to obtain explicit formulas. In that Box Model, the
intensity function $\DocRateFn_\Doc$ of request for any document
$\Doc$ is piecewise constant, that is, $\DocRateFn_\Doc(s) =
\lambda_\Doc \cdot \II{ \CatArr_\Doc \leq s \leq \CatArr_\Doc +
  \DocSpan_\Doc }$ where we independently choose the pair of
(possibly dependent) random variables $\lambda_\Doc$ and
$\DocSpan_\Doc$ from the catalog arrival process $\CatPP$;
(\ref{ndbar}) now reduces to $\overline{\NReqs}_\Doc =
\lambda_\Doc \DocSpan_\Doc$.  

Please note that many choices of pattern for $\DocRateFn_\Doc$
are possible. In particular, it is possible to choose intensity
functions with unbounded support, such as the exponential
function (in this case, the lifespan of a document could be
defined for example as the time interval for 90\% of the document
requests to occur). Our choice of a piecewise
intensity function is 
consistent with Insight I3 
gained from Section~\ref{sec:semiExperiments}.

\begin{pro}
\label{pro:DDMean_box_model}
For the Box Model, the mean of the process $\DD^0$ is given by
\begin{align*}
\DDMeanFn(t) & = 
\CatArrRate \, 
\ESS{2t + \left(1 - e^{-\lambda t} \right)\left(\DocSpan - t - \frac{2}{\lambda} \right)}{\DocSpan \geq t} \\
 & + \CatArrRate \, 
 \ESS{2\DocSpan + \left(1 - e^{-\lambda \DocSpan} \right)\left(t - \DocSpan - \frac{2}{\lambda} \right)}{\DocSpan < t}
\end{align*}
for all $t \geq 0$, where the pair of positive variables $(\lambda,\tau)$ is  distributed as any pair $(\lambda_\Doc,\tau_\Doc)$.
\end{pro}

To interpret the latter expression of $\DDMeanFn(t)$, assume $\DocSpan_\Doc = \DocSpan_0$ and $\DocRateFn_\Doc = \lambda_0$ are constants; then $\DDMeanFn(t)$ grows non-linearly in $t$ if $ t < \DocSpan_0$ and linearly otherwise; in the latter case, we can write the mean function as 
$\DDMeanFn(t) = 
\DDMeanFn(\DocSpan_0) + \CatArrRate(t - \DocSpan_0)(1-e^{-\lambda_0 \DocSpan_0 })$ 
which is just the mean number of new objects up to time $\DocSpan_0$, plus the
mean number of arrivals to the catalog in interval $[\DocSpan_0, t]$ penalized
by the probability that the document has at least one request. 

To finally specify the Che approximation for the Box Model, we now state the following.
\begin{pro}
\label{pro:che_approx_box_model}
Under the Che approximation and for the Box Model, the
conditional expectation of the expected number of hits to
document $\Doc$ is given by
\[
\MeanHitsDoc =
\begin{cases}
\lambda_\Doc \DocSpan_\Doc - 1 + e^{- \lambda_\Doc \DocSpan_\Doc} 
& \text{if} \; \DocSpan_\Doc < \CharTime{\Csize}, 
\\ \\
(\lambda_\Doc \DocSpan_\Doc - 1)(1 - e^{-\lambda_\Doc \CharTime{\Csize}}) 
 + \lambda_\Doc \CharTime{\Csize} e^{-\lambda_\Doc \CharTime{\Csize}}
& \text{if} \; \DocSpan_\Doc \geq \CharTime{\Csize}.
\end{cases}
\]
\end{pro}
We then conclude from Proposition~\ref{pro:che_approx_box_model}
that the expected number of hits to all documents is given by
\begin{align}
& \E{H} = \int \int_{\lambda > 0, \tau < t_c} \left [ \lambda \DocSpan - 1 + e^{- \lambda \DocSpan} \right ] f(\lambda,\tau)\mathrm{d}\lambda\mathrm{d}\tau \; +
\label{EHBox}
\\
& \; \int \int_{\lambda > 0, \tau \geq t_c} \left [ (\lambda \DocSpan - 1)(1 - e^{-\lambda \CharTime{\Csize}}) 
 + \lambda \CharTime{\Csize} e^{-\lambda \CharTime{\Csize}} \right ] f(\lambda,\tau)\mathrm{d}\lambda\mathrm{d}\tau
 \nonumber
\end{align}
where $f$ denotes the joint probability density of the pair $(\lambda,\tau)$,
and with $t_C$ derived by (\ref{tC}) via the expression of
$\DDMeanFn(t)$ obtained in Proposition \ref{pro:DDMean_box_model}.


\section{Model Validation}
\label{sec:ME}


The aim of this final section is to assess the validity of our
Box model for the calculation of the hit ratio (as derived in
Proposition~\ref{pro:che_approx_box_model}), when compared to the
values obtained by a direct simulation. To this end, we first
detail the computation of the necessary statistics to use our
model, namely \emph{\textbf{(i)}} the catalog arrival intensity
$\gamma$, \emph{\textbf{(ii)}} the mean $\DDMeanFn(t)$ for all $t
\geq 0$, and \hbox{\emph{\textbf{(iii)}} the} hit ratio $\mathrm{HR} =
(\sum_{\Doc} H_d) / (\sum_{\Doc} n_d) = \E{\HitsDoc}/\E{\NReqs_\Doc}$.


\begin{figure*}[h!t]
		\begin{subfigure}[b]{0.45\textwidth}
		\centering
		\input{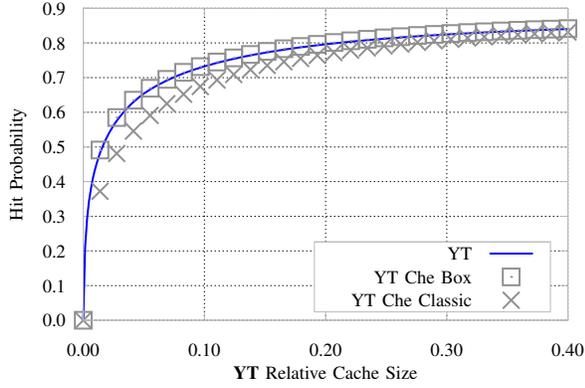}
		\label{fig:otu_hr_vs_est}
		\caption{\it  YT trace.}
		\end{subfigure}%
		\quad\quad
		\begin{subfigure}[b]{0.45\textwidth}
		\centering
		\input{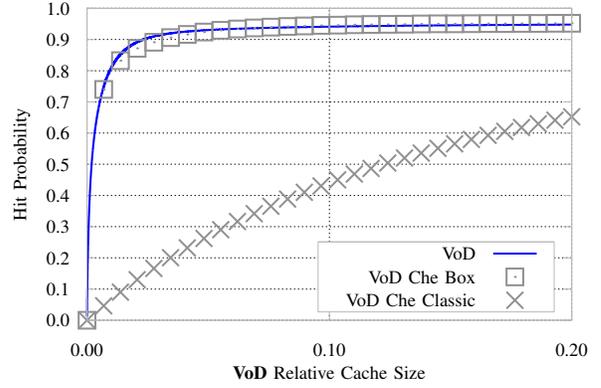}
		\label{fig:vod_hr_vs_est}
		\caption{\it VoD trace.}
		\end{subfigure}%
		\caption{\it Fittings for the Che estimation}
		\label{fig:hr_che}
\end{figure*}

\textit{(i)} Let first $\SNobj$ be the number of documents in our sample and denote by $\WinSize$
the size of the observation window; we can then estimate the catalog arrival
rate $\CatArrRate$ by $ \Est{\CatArrRate} = \SNobj/\WinSize$;

\textit{(ii)} We have noted in Section \ref{sec:est_quant} that estimators
$\Est{\DocSpan}$ and $\Est{\DocReqRate}$, respectively expressed in (\ref{EstimTau}) and (\ref{EstimLambda}), are not available for
documents with only one request. This sub-sample has, however, a
considerable size (58\% of all documents) and cannot be
neglected in a direct application of Proposition \ref{pro:DDMean_box_model}. 

To incorporate this data, we use the approximation
discussed in~\cite{KauffmannITC2013} where the set of documents
requested only once is represented by a ``noise'' process. Let $\DDMeanFn_1$ (resp. $\DDMeanFn_2$) denote the mean function of that noise process (resp. the mean function associated with the ``non-noise'' part of the process), with $\DDMeanFn = \DDMeanFn_1 + \DDMeanFn_2$. We can separate the noise process from the rest of the request process and, using a procedure similar to that of
Proposition~\ref{pro:DDMean_box_model}, we easily obtain an
explicit formula for $\DDMeanFn_1(t)$ (omitted here for brevity); 
the latter together with the formula for $\DDMeanFn(t)$ in Proposition~\ref{pro:DDMean_box_model} then gives
$
\DDMeanFn_2(t) = \DDMeanFn(t) - \DDMeanFn_1(t) = \CatArrRate \,
\E{L(\DocReqRate, \DocSpan, t)} \text{, where}
\label{eq:Xsi2estimate}
$
\begin{align*}
&L(\DocReqRate, \DocSpan, t) = 
\\
& \left [ 2t (1-e^{\DocReqRate t } ) + \left(1 - e^{\DocReqRate t } - \DocReqRate t e^{\DocReqRate t} \right) \left(\DocSpan - t - \frac{4}{\DocReqRate} \right) \right ] \II{\DocSpan \geq t} \; + \\
& 
 \left [ 2\DocSpan (1-e^{\DocReqRate\DocSpan} ) + \left(1 - e^{\DocReqRate\DocSpan} - \DocReqRate\DocSpan e^{\DocReqRate\DocSpan} \right) \left(t - \DocSpan - \frac{4}{\DocReqRate} \right) \right ] \II{\DocSpan < t}.
\end{align*}
Recall that the documents for which we have an estimate of the
pair $(\DocReqRate_\Doc, \DocSpan_\Doc)$ are precisely those
accounted into $\DDMeanFn_2$. We thus estimate $\Est{\DDMeanFn}_2(t) = 
\Est{\CatArrRate} \times \E{L(\DocReqRate, \DocSpan, t)}$, with
the expectation taken w.r.t. the empirical distribution of
$\left(\lambda,\tau\right)$ in the trace. 

Finally, let $\SNobj_1$ and $\SNobj_2$ be the number of documents with one, and more than one requests, respectively. We then estimate $\DDMeanFn_1(t)$ by the mean function of a
homogeneous Poisson process, that is, $\Est{\DDMeanFn}_1(t) =
\SNobj_1 \times t/\WinSize$. 
The estimator of the characteristic time associated with the Che
approximation is therefore given by
$\Est{t}_\Csize= \Est{\DDMeanFn}^{-1}(\Csize)$, with 
$\Est{\DDMeanFn}(t) = \Est{\DDMeanFn}_1(t) + \Est{\DDMeanFn}_2(t)$.

\textit{(iii)} Concerning the hit ratio HR, we must similarly take the documents with just one request into account. Note that since the documents pertaining to the noise process do not produce hits, we can write
\[
\mathrm{HR} = \frac{\E{\HitsDoc}}{\E{\NReqs_\Doc}} = 
\frac{\ESS{\HitsDoc}{\NReqs_\Doc \geq 2}}
{\ESS{\NReqs_\Doc}{\NReqs_\Doc \geq 1}} 
= \frac{
\EC{\ES{\HitsDoc \II{\NReqs_\Doc \geq 2}}{\Doc}}
{\NReqs_\Doc \geq 2}}
{\EC{\NReqs_\Doc}{\NReqs_\Doc \geq 2} + \displaystyle \frac{\PP{\NReqs_\Doc = 1}}{\PP{\NReqs_\Doc \geq 2}}}.
\]
Let $\MeanHitsDoc = M(\DocSpan_\Doc, \DocReqRate_\Doc,
\CharTime{\Csize})$ be the conditional expectation of the number
of hits given by Proposition~\ref{pro:che_approx_box_model};
by similar arguments to those used in \emph{(ii)}, the
numerator of the hit ratio is thus estimated by
\begin{align*}
\EC{\MeanHitsDoc}{\NReqs_\Doc \geq 2}
& \approx
\E{M(\DocSpan, \DocReqRate, \Est{t}_\Csize)}
\end{align*}
with
the expectation taken w.r.t. the empirical distribution of
$\left(\lambda,\tau\right)$ in the trace. 
As to the term $\EC{\NReqs_\Doc}{\NReqs_\Doc \geq 2}$, it can be
computed as the average 
number of requests in the corresponding sub-sample. Finally, the ratio
$\PP{\NReqs_\Doc = 1}/\PP{\NReqs_\Doc \geq 2}$ in the denominator
is estimated by 
$\PP{\NReqs_\Doc = 1}/\PP{\NReqs_\Doc \geq 2} \approx \SNobj_1/\SNobj_2$.

Using the above estimators, we can eventually compare the hit
ratio derived from the Box Model to that obtained by simulation
for each trace, as depicted in Figure~\ref{fig:hr_che}. For
comparison purpose, we provide also the estimation of the hit
ratio obtained by the Che approximation when the request process
is assumed to be IRM. 
For the YouTube traffic, the Box Model improves the accuracy by
one order of magnitude compared to the estimation with an IRM
process, with respective MARE of 0.5\% and 4.1\%. For the VoD
traffic, the improvement is even more spectacular, due to the
large duration of the trace. The IRM is far from estimating
properly the hit ratio with a MARE of 17.2\% (this value is
significantly decreased by including the tail of the curve, not
plotted here, and where the IRM converges towards the correct
value). On the other hand, the Box Model estimates accurately the hit
ratio, with a MARE of 0.6\%. This validates our model.



\section{Conclusion}
\label{sec:conclusion}


The current literature on the performance of caching systems ignores the fact that
content is produced and becomes out-dated everyday. The
consequence for caches is non-negligible, as requests for a given
document are concentrated within its lifetime and the request process
is thus non stationary. 
This paper addresses the issue of catalog dynamics.
Based on two traffic traces, we provide evidence for the impact of the
catalog dynamics and 
identify the core structures of the request process.
We then propose a general model for the aggregated
request process and  provide an estimate of the hit
ratio of a LRU cache fed by such a request process. 

Our results show that the document request process can be easily
described, as far as caching is concerned, in terms of basic
document statistics: the document lifespan and its average
request intensity (within its lifespan). As expected, the hit
ratio is mainly driven by the distribution
of the number of requests for each document. The distribution of
request intensities, however, has a secondary impact on the hit
ratio: higher intensities (and thus shorter lifespan) lead to
higher performance, which confirms basic intuition. 

Our proposed model currently uses documents as a basic unit. In
practice, however,  
bandwidth and cache size are counted in bytes. Additionally, in the
case of video streaming, the downloads of videos are frequently
interrupted because users switch to another video. As a further
study, we intend to account for both the size distribution of videos and the impact of these interruptions on caching. Our model for the catalog dynamics can also be
improved: the catalog size in our model is stationary, while it
actually increases with time. 



%
\bibliographystyle{abbrv}
\bibliography{dynamic_catalog_arxiv}  
%
%


\appendix


\section{Software Tools}
In Section \ref{sec:est_quant}, we calculate the unidimensional kernel densities for variables $\log \lambda$ and $\DocSpan$ by means of the \texttt{density} function provided by the statistical software R~\cite{Rsoft}. For the bi-dimensional kernel density estimation, we use the \texttt{kde2d} function of the R package \texttt{MASS}~\cite{MASS}.

\section{Technical Proofs}
In order to justify Proposition \ref{pro:DD_laws}, we assert some technical
properties for the processes involved in the modeling. The following notation will be used throughout:

- given a measurable space $S$, $\delta_x$ denotes the Dirac mass at point $x \in S$;

- for any function $f:S \rightarrow \RR^+$ on $S$ equipped with measure $\mu$, we set $\langle f , \mu \rangle = \int_S f \mathrm{d} \mu$;

- for any point process $\xi$ on space $S$ with associated intensity measure
$\mu$, the Laplace functional $\LF_\xi$ of $\xi$ is
defined by $\LF_\xi(f) = \mathbb{E}[e^{- \langle \mu , f \rangle}] = \mathbb{E}[\exp(-\sum_{a
\in \xi} f(a)]$ for any measurable function $f:S \rightarrow \RR^+$. 

We finally recall (\cite{KALL}, Lemma 12.2) the following properties of Laplace
functionals for Poisson and general Marked Point processes.
\begin{property}
\begin{itemize}
\item[(i)] Point process $\xi$ is Poisson if and only if $\LF_\xi(f) = \exp(-\langle \mu, 1 - e^{-f} \rangle)$ for any measurable positive measurable function $f$; 
\item[(ii)] let $\xi$ be an arbitrary point process; for any $a \in S$, let $Z_a$ be a random variable with values in a space $M$ and whose probability distribution is denoted by $\nu(a,\cdot)$; we assume that all $Z_a$, $a \in S$, are mutually independent. Consider the marked point process $\widetilde{\xi} = \sum_{a \in \xi} \delta_{(a,Z_a)}$. For any measurable positive function $\widetilde{f}: S \times M \rightarrow \RR^+$, we set 
$$
f(a) = - \log \int_M e^{-\widetilde{f}(a,z)} \nu(a,\mathrm{d}z), \; \; a \in S;
$$
then the equality $\LF_{\widetilde{\xi}}(\widetilde{f}) = \LF_\xi(f)$ holds. In particular, if $\xi$ is also Poisson, then
\begin{equation}
\mathcal{L}_{\widetilde{\xi}}(\widetilde{f}) = \exp \left \{ - \int_S \mathbb{E}[1 - e^{-\widetilde{f}(a,Z_a)}] \mathrm{d}\mu(a) \right \}
\label{LaplafunctMarked}
\end{equation}
for any measurable positive function $\widetilde{f}: S \times M \rightarrow \RR^+$.
\end{itemize}
\label{PropPoint}
\end{property}

We first show a decomposition property for marked Poisson point processes.

\begin{lem}[Splitting by marks]
Let $\PPr$ denote a Poisson point process on the real line with intensity measure $\PPrMM$. Given a measurable space $M$, define the marked Poisson process $\MPPr$ on $\RR \times M$
by 
\[
\MPPr = \sum_{a \in \PPr} \delta_{(a, \RV_a)}
\]
with independent (but not necessarily identically distributed) random variables $\RV_a$ with values in $M$. If $\FSeq{M}{n}$ is a partition of $M$, then each process
\[
\MPPr_k = \sum_{a \in \PPr} \delta_{(a, \RV_a)} \cdot \II{\RV_a \in M_k}, \; \; 1 \leq k \leq n,
\]
is a marked Poisson point process with intensity measure $\mu_k$ defined by 
\[
\mu_k (\mathrm{d}a \times B) = \PP{\RV_a \in M_k} \PC{\RV_a \in B}{\RV_a \in M_k} \mathrm{d}\PPrMM(a)
\] 
for any measurable set $B \subset M_k$. Processes $\MPPr_k$, $1 \leq k \leq n$, are also mutually independent.
\label{LemmaSplit}
\end{lem}
\begin{IEEEproof} Let $\FSeq{\widetilde{f}}{n}$ be positive functions defined on $\RR \times M$; the Laplace functional $\LF$ of the tuple $(\FSeq{\MPPr}{n})$ is then given by
\[
\!\!\!\!\! \!\!\!\!\! \LF(\widetilde{f}_1,...,\widetilde{f}_n) = \E{\exp{\left\{ -\sum_{k=1}^n \langle \MPPr_k , \widetilde{f}_k \rangle \right\}}}
=
\E{\exp{ \left\{- \langle \MPPr , \widetilde{f} \rangle \right\}}}
\]
where 
\[
\widetilde{f}(a, x) = \sum_{k=1}^n \widetilde{f}_k(a, x) \II{x \in M_k}.
\]
Applying then (\ref{LaplafunctMarked}) to ground Poisson process $\Gamma$, we successively obtain
\begin{multline*}
 \LF(\widetilde{f}_1,...,\widetilde{f}_n) = \exp{\left\{ - \int_{\RR}  \E{1 - e^{-\widetilde{f}(a, Z_a)} } \mathrm{d}\PPrMM(a) \right\} } =
\\
 \exp{\left\{ - \int_{\RR}  \sum_{k=1}^n \ESS{1 - e^{-\widetilde{f}_k(a, Z_a)} }{Z_a \in M_k} \mathrm{d}\PPrMM(a) \right\} } = 
\\
 \prod_{k=1}^n \exp{\left\{ - \int_{\RR} \ESS{1 - e^{-\widetilde{f}_k(a, Z_a)} }{Z_a \in M_k} \mathrm{d}\PPrMM(a) \right\} } = 
\\
\!\!\!\!\!\!\!\!\! \prod_{k=1}^n \exp{\left\{ - \int_{\RR} \EC{1 - e^{-\widetilde{f}_k(a, Z_a)} }{Z_a \in M_k}  \PP{Z_a \in M_k} \mathrm{d}\PPrMM(a) \right\} }
\end{multline*}
which, by  Property \ref{PropPoint}.$(i)$, concludes the proof.
\end{IEEEproof}

\begin{IEEEproof}[Proof of Proposition \ref{pro:DD_laws}]
We now prove Proposition \ref{pro:DD_laws} in two steps:
\begin{enumerate}
\item
$\DD^0(t)$ is a Poisson Process with mean
function $\DDMeanFn(t)$;
\item
Processes $(\DD^s_t)_{t \geq s}$ and $(\DD^0_{t - s})_{t \geq 0}$
have identical distributions. 
\end{enumerate}
\paragraph{First step}
Given the ground process $\Gamma$, define the marked ground process $\MarkedCatPP$ by
$$
\MarkedCatPP = \sum_d \delta_{a_d,\DocReqPP_d}
$$
where the mark $\DocReqPP_\Doc$ is the request arrival process for document $\Doc$ introduced in Section \ref{CARP}. Let then $\FRPP^s$ denote the point process defined by the sequence of first request times in $[s,+\infty)$ (if they exist) for any document. Starting from process $\MarkedCatPP$, we will construct $\FRPP^s$ in two steps: 

- first, consider the space $M_1$ of point processes $\DocReqPP$ which have at least one point in interval $[s,+\infty)$, that is, 
$$
M_1 = \{ \DocReqPP \; \vert \; \DocReqPP [s,+\infty) \geq 1\},
$$
its complement $M_2 = \{ \DocReqPP \; \vert \; \DocReqPP [s,+\infty) = 0\}$ and the union $M = M_1 \cup M_2$; referring then to Lemma \ref{LemmaSplit}, we can define marked point processes $\MPPr_1$ and $\MPPr_2$. By Lemma \ref{LemmaSplit}, in particular, the intensity measure $\mu_1$ of $\MPPr_1$ is given by 
\begin{equation}
\mu_1(\mathrm{d}a \times B) = 
\PP{ \DocReqPP_\Doc \in B, \; \DocReqPP_\Doc[s, +\infty) \geq 1}  \times \CatArrRate \mathrm{d}\CatArr
\label{mu1}
\end{equation}
for any $B \subset M_1$, where $a$ denotes $a_d$ for short;

- secondly, $\FRPP^s$ can be written in terms of of $\MarkedCatPP$ as
\[
\FRPP^s = \sum_{\CatArr \in \MarkedCatPP_1} \delta_{T(\DocReqPP_\Doc)}
\]
where $T(\DocReqPP_\Doc)$ denotes the first point of $\DocReqPP_\Doc$ in $[s, +\infty)$ (note that we have split process $\MarkedCatPP$ in such a way that $T$ is well-defined in the subset $M_1$). 

To show that process $\FRPP^s$ is a Poisson point process, we now calculate the exponent of its Laplace functional: let $f$ be a positive function, we have
\begin{align}
\expLF{f} &= -\log{\E{e^{-\FRPP^sf}}} = 
-\log{\E{\exp{
	\left\{
		-\sum_{\CatArr \in \MarkedCatPP_1} f(T(\DocReqPP_\Doc))
	\right\}
}}} 
\nonumber \\
&=
-\log{\E{e^{- \MarkedCatPP_1 \widehat{f}}}}
\label{Gamma1Tilde}
\end{align}
where $\widetilde{f}(\CatArr, \DocReqPP_\Doc) =
f(T(\DocReqPP_\Doc))$. We are therefore left to calculate the
Laplace functional of process $\MarkedCatPP_1$ in order to obtain
that of process $\FRPP^s$. As the marked Poisson process
$\MarkedCatPP_1$ has the intensity measure $\mu_1$ given in
(\ref{mu1}), formula (\ref{LaplafunctMarked}) enables us to
further write expression(\ref{Gamma1Tilde}) as
\begin{align*}
& \expLF{f} =
\\
& 
	\CatArrRate \int_{\RR} \PP{\DocReqPP_\Doc[s, +\infty) \geq 1} 
	\EC{1 - e^{-f(T(\DocReqPP_\Doc))}}{\DocReqPP_\Doc[s, +\infty) \geq 1}
	\mathrm{d}\CatArr
\\
&	= \CatArrRate \int_{\RR}
	\ESS{1 - e^{-f(T(\DocReqPP_\Doc))}}
			{\DocReqPP_\Doc[s, +\infty) \geq 1}
	\mathrm{d}\CatArr
\\
&	= \CatArrRate \int_{\RR}
	\sum_{k=1}^{+\infty}
	\ESS{1 - e^{-f(T(\DocReqPP_\Doc))}}
		 {\DocReqPP_\Doc[s, +\infty) = k}
	\mathrm{d}\CatArr.
\end{align*}
Recall that, given $\DocRateFn_\Doc$, the process $\DocReqPP_\Doc$ is
Poisson with intensity function $\DocRateFn_\Doc$, and thus $\DocReqPP_\Doc[s, +\infty)$ is a Poisson random variable with parameter
$$
m_\Doc(s) = \int_s^{+\infty} \Lambda_d(u) \mathrm{d}u
$$
(note our assumptions ensure that $m_\Doc(s)$ is almost surely finite). By conditioning on $\DocRateFn_\Doc$, the exponent $\expLF{f}$ above can be further expressed in terms of $\overline{\NReqs}_\Doc$ as
\begin{align} 
\label{eq:laplace_funct_full_conditioned}
\expLF{f} = \CatArrRate \int_{\RR} \mathbb{E} & \left [
\sum_{k=1}^{+\infty}
\ECSS{1 - e^{-f(T(\DocReqPP_\Doc))}}
	 {\DocRateFn_\Doc}
	 {\DocReqPP_\Doc[s, +\infty) = k} \right.
\nonumber \\
& \left.
	 \times \; e^{-m_\Doc(s)} \frac{m_\Doc(s)^k}{k!} \right ] \mathrm{d}\CatArr.
\end{align}
Now, given the event $\DocReqPP_\Doc[s,+\infty) = k$, the distribution
of the request arrival times follows that of a $k$-sample with density
\[
\dens_\Doc(u) = \rec{m_\Doc(s)} \DocRateFn_\Doc(u) 
\cdot
\I_{[s \maxi a, \CatArr + \DocSpan_\Doc]}(u), \; \; u \in \RR;
\]
we also denote by $\dist_\Doc$ the associated \cdf of $\dens_\Doc$. Given all
the conditionings by $\Lambda_\Doc$ and $\DocReqPP_\Doc[s,+\infty) = k$, $T(\DocReqPP_\Doc)$ is distributed as the minimum of a $k$-sample drawn from distribution $\dist_\CatArr$; we consequently have
\begin{align*}
& \ECSS{1 - e^{-f(T(\DocReqPP_\Doc))}}
	 {\DocRateFn_\Doc}
	 {\DocReqPP_\Doc[s, +\infty) = k} \, = 
\\
& \int_{\RR}
(1 - e^{-f(u)}) k \,
\dens_\Doc(u) \, (1 - \dist_\Doc(u))^{k-1} \, \mathrm{d}u.
\end{align*}
Noting that
\[
\sum_{k=1}^{+\infty}
k \, (1 - \dist_\Doc(u))^{k - 1} e^{-m_\Doc(s)} \frac{m_\Doc(s)^k}{k!}
=
m_\Doc(s) e^{-m_\Doc(s) \dist_\Doc(u)},
\]
expression (\ref{eq:laplace_funct_full_conditioned}) reduces to
\begin{multline*}
\! \! \! \! \! \! \! \expLF{f} = 
	\CatArrRate \int_\RR
	\E{
	\int_\RR
	(1 - e^{-f(u)})
	m_\Doc(s) \, \dens_\Doc(u)
	e^{- m_\Doc(s) \dist_\Doc(u)}
	\, \mathrm{d}u
	}
	\mathrm{d}\CatArr 
\\ = 
\CatArrRate \, \E{
	\int_\RR
	\int_\RR
	(1 - e^{-f(u)})
	m_\Doc(s) \, \dens_\Doc(u)
	e^{- m_\Doc(s) \dist_\Doc(u)}
	\, \mathrm{d}\CatArr 
	\mathrm{d}u
}
\\ = 
\CatArrRate \int_\RR
(1 - e^{-f(u)}) \,
\E{
	\int_\RR
	m_\Doc(s) \, \dens_\Doc(u)
	e^{- m_\Doc(s) \dist_\Doc(u)}
	\, \mathrm{d}\CatArr 
}
\mathrm{d}u
\end{multline*}
Noting the identity of sets
\[
\left\{ 
	(\CatArr,u): s \maxi \CatArr \leq u \leq \CatArr + \DocSpan 
\right\}
=
\left\{
	(\CatArr,u):  u - \DocSpan \leq \CatArr \leq u, \, u \geq s
\right\}
\]
in the definition of density $\dens_\Doc$ and distribution function $\dist_\Doc$, the latter equation for $\expLF{f}$ reads
\begin{align*}
\expLF{f} = & \,
\CatArrRate \int_s^\infty
(1 - e^{-f(u)}) \,
\\
& \; \times \E{
	\int_{u -\DocSpan}^u
	\DocRateFn_\CatArr(u)
	\exp{
		\left(- 
		\int_s^u
		\DocRateFn_\CatArr(v) 
		\, \mathrm{d}v 
		\right)}
	\, \mathrm{d}\CatArr 
}
\, \mathrm{d}u
\end{align*}
We therefore conclude that $\FRPP^s$ is a Poisson point process with intensity measure
\[
\CatArrRate \,
\E{ \int_{u - \DocSpan}^u \DocRateFn_\Doc(u)
	\exp{
		\left(- 
		\int_s^u \DocRateFn_\Doc(v) \, \mathrm{d}v 
		\right)}
	\, \mathrm{d}\CatArr 
}
\, \mathrm{d}u, \; \; u \geq s,
\]
and the mean function of $\DD^s$ for $t \geq s$ is then given by
\begin{multline*}
\DDMeanFn^s(t) = 
\CatArrRate \int_s^t  
\E{ \int_{u - \DocSpan}^u \DocRateFn_\Doc(u)
	\exp{
		\left(- 
		\int_s^u \DocRateFn_\Doc(v) \, \mathrm{d}v 
		\right)}
	\, \mathrm{d}\CatArr 
}
\, \mathrm{d}u \; 
\\  
= \CatArrRate \, 
\E{
\, \int_\RR
\, \int_{s \maxi \CatArr}^{t \mini \CatArr + \DocSpan}
 \DocRateFn_\Doc(u)
	\exp{
		\left\{- 
		\int_s^u \DocRateFn_\Doc(v) \, dv 
		\right\}}
	\, \mathrm{d}u 
	\, \mathrm{d}\CatArr 
} \; 
\\
\!\!\!\!\!\!\!\!\!\!\!\!\!\!\!\!\!\!\!\! = 
 \CatArrRate  \,
\E{
\, \int_\RR
\exp{\left\{ - \int_s^{s \maxi \CatArr}  \DocRateFn_\CatArr(v)\, \mathrm{d}v  \right\}} 
- \exp{\left\{ - \int_s^{t \mini \CatArr + \DocSpan}  \DocRateFn_\CatArr(v)\, \mathrm{d}v  \right\}}
	\, \mathrm{d}\CatArr
}.
\end{multline*}
To further simplify the latter expression, we note that
\[
\int_s^{s \maxi \CatArr} \DocRateFn_\Doc(v) \, \mathrm{d}v = 0
\]
since if $s < \CatArr$, then $\DocRateFn_\Doc$ is zero in $[s, \CatArr]$; 
the above  expression for $\DDMeanFn^s(t)$ therefore reduces to
\begin{align*}
\DDMeanFn^s(t) & =
\CatArrRate \, \E{
\, \int_\RR \left ( 1 - \exp{\left\{ - \int_{s \maxi \CatArr}^{t \mini \CatArr + \DocSpan}  
\DocRateFn_\Doc(v)\, \mathrm{d}v  \right\}} \right ) \, \mathrm{d}\CatArr 
}
\\ 
& =
\CatArrRate \,
\E{
\, \int_{-\infty}^t \left ( 1 -\exp{\left\{ - \int_s^t \DocRateFn_\Doc(v)\, \mathrm{d}v  \right\}} \right )
\, \mathrm{d}\CatArr 
}
\end{align*}
where the last equality holds because $\DocRateFn_\Doc$ is zero out of interval 
$[\CatArr, \CatArr + \DocSpan]$. Applying the latter expression
to $s =0$ readily gives the claimed formula for $\DDMeanFn^0(t)
= \DDMeanFn(t)$, $t \geq 0$, which concludes the first step.

\paragraph{Second step} For $s_1 < s_2$, let  $\DDMeanFn^{s_1}$ and $\DDMeanFn^{s_2}$ be the mean functions of $\DD^{s_1}$ and $\DD^{s_2}$, respectively. As a Poisson process is completely determined by its mean 
function, it is sufficient to prove that 
\[
\DDMeanFn^{s_1}(t - (s_2 - s_1)) = \DDMeanFn^{s_2}(t)
\]
for all $t$. Setting $\Delta s = s_2 - s_1$, we thus have
$$
\DDMeanFn^{s_1}(t - \Delta s) 
= 
\CatArrRate  \E{\int_{-\infty}^{t - \Delta s} \! 
\left (\! 1 - \exp{\left\{ 
- \! \int_{s_1}^{t  -\Delta s} \DocRateFn_\Doc(v) \mathrm{d}v  
\right\}} \right) \mathrm{d}\CatArr } 
$$
where $a$ stands for $a_d$ (the document $d$ arrival time) for short; using the variable change $a' = a - \Delta s$, the intensity function $v \mapsto \Lambda_d(v)$ with support $[a_d,a_d+\tau_d] = [a,a+\tau]$ changes to the translated function $v \mapsto \Lambda_d(v + \Delta s)$ with support $[a-\Delta s, a + \tau - \Delta s]$. We then obtain
\begin{align*}
& \DDMeanFn^{s_1}(t - \Delta s) \; =
\\
& \CatArrRate \, \mathbb{E} \left [ \int_{-\infty}^{t} \left ( 1 -
\exp \left\{ - \int_{s_1}^{t  - \Delta s} 
\DocRateFn_{\Doc}(v +  \Delta s) \, \mathrm{d}v  \right\} \right ) \, \mathrm{d} \CatArr' \right ]
 \; =
\\
& \CatArrRate \, \mathbb{E} \left [ \int_{-\infty}^{t} \left ( 1 - \exp \left\{ - \int_{s_2}^{t} 
\DocRateFn_{\Doc}(v) \, \mathrm{d} v  \right\} \right ) \, \mathrm{d} \CatArr' \right ]
  \; = \; \DDMeanFn^{s_2}(t)
\end{align*}
which concludes the proof.
\end{IEEEproof}

We can now deduce the explicit expression for $\DDMeanFn(t)$ for Box Model.

\begin{IEEEproof}[Proof of Proposition \ref{pro:DDMean_box_model}] 
For brevity we denote $\DocReqRate_\Doc = \DocReqRate$ and $\DocSpan_\Doc = \DocSpan$. This presents
no ambiguity since the distribution of the pair $(\DocReqRate_\Doc, \DocSpan_\Doc)$ does not depend on
the document $\Doc$.
By Proposition \ref{pro:DD_laws}, we derive that
\begin{align*}
\DDMeanFn(t) & =  \CatArrRate \, \E{ \, \int_{-\infty}^t 1 - e^{- 
\DocReqRate 
(t \mini (\CatArr + \DocSpan) - \CatArr^+)^+} \, \mathrm{d}\CatArr }
\\ 
& = \CatArrRate \, \E{ \, \int_{-\DocSpan}^t 1 - e^{ - \DocReqRate (t \mini (\CatArr + \DocSpan) - \CatArr^+) } \, \mathrm{d}\CatArr }
\\ 
& = \CatArrRate \, \E{ t + \DocSpan } - \CatArrRate \, \E{ \int_{-\DocSpan}^t e^{- \DocReqRate (t \mini (\CatArr + \DocSpan) - \CatArr^+) } \, \mathrm{d}\CatArr }.
\end{align*}
On the event $(\DocSpan < t)$, the inner integral becomes
\begin{align*}
& \int_{-\DocSpan}^t e^{ - \DocReqRate (t \mini (\CatArr + \DocSpan) - \CatArr^+) }\, \mathrm{d}\CatArr \; =
\\
& \int_{-\DocSpan}^0 e^{ - \DocReqRate (\CatArr + \DocSpan)} \, d\CatArr + 
\int_0^{t - \DocSpan} e^{ - \DocReqRate \DocSpan } \, \mathrm{d}\CatArr + \int_{t - \DocSpan}^t
e^{ - \DocReqRate (t - \CatArr) }
\, \mathrm{d}\CatArr \; =
\\
& \frac{1 - e^{-\DocReqRate \DocSpan}}{\DocReqRate}
+ e^{- \DocReqRate \DocSpan}(t - \DocSpan)
+ \frac{1 - e^{-\DocReqRate \DocSpan}}{\DocReqRate} =
\\ 
& \DocSpan - t
- (1 - e^{-\DocReqRate \DocSpan})
\left(t - \DocSpan - \frac{2}{\DocReqRate} \right).
\end{align*}
On the event $(\DocSpan \geq t)$, we similarly obtain
\begin{align*}
&\int_{-\DocSpan}^t e^{ - \DocReqRate (t \mini (\CatArr + \DocSpan) - \CatArr^+) } \, \mathrm{d}\CatArr \; =
\\
& \int_{-\DocSpan}^{t - \DocSpan} e^{ - \DocReqRate (\CatArr + \DocSpan) }\, d\CatArr
+ \int_{t - \DocSpan}^0 e^{ - \DocReqRate t}\, \mathrm{d}\CatArr
+ \int_0^t e^{ - \DocReqRate (t - \CatArr)}\, \mathrm{d}\CatArr \; =
\\
& \frac{1 - e^{-\DocReqRate t}}{\DocReqRate} + e^{-\DocReqRate t}(\DocSpan - t)
+ \frac{1 - e^{-\DocReqRate t}}{\DocReqRate} \; =
\\
& t - \DocSpan
- (1 - e^{-\DocReqRate t})
\left(\DocSpan - t - \frac{2}{\DocReqRate} \right)
\end{align*}
from which the final formula for $\DDMeanFn(t)$ follows.
\end{IEEEproof}

\begin{IEEEproof}[Proof of Proposition \ref{pro:che_approx_box_model}]
As the request process $\DocReqPP_\Doc$ for any document $\Doc$ is Poisson
then, given $\DocSpan_\Doc$ and the event $\{\NReqs_\Doc = k\}$, the tuple
$(\FSeq{\Req}{k})$ follow the distribution of the order statistics of $k$
uniform random variables over $[0, \DocSpan_\Doc]$. Consequently, the variable
$(\Req_\iReq - \Req_{\iReq - 1})/\DocSpan_\Doc$ follows a $\BetaD{1}{k}$
distribution.

In the event $(\DocSpan_\Doc < \CharTime{\Csize})$, every request is a hit except for
the first request, regardless of the values of $\FSeq{\Req}{k}$; on this event, we thus derive from (\ref{eq:EH_unfoldedBIS}) that
\begin{align*}
& \MeanHitsDoc \; = 
\\
& \sum_{k = 2}^{+\infty}
\left(
	\sum_{\iReq = 2}^{k} 
	\ECSS{\II{\Req_\iReq - \Req_{\iReq - 1} < t_C}}
	     {\Doc}{\NReqs_\Doc = k}
\right)
e^{-\DocReqRate \DocSpan} 
\frac{\left( \DocReqRate \DocSpan \right)^k}{k!} =
\\
& \sum_{k = 2}^{+\infty} (k-1) e^{-\DocReqRate \DocSpan} \frac{\left( \DocReqRate \DocSpan \right)^k}{k!} = \DocReqRate \DocSpan - 1 + e^{- \DocReqRate \DocSpan}.
\end{align*}
where we set again $\DocReqRate_\Doc = \DocReqRate$ and $\DocSpan_\Doc = \DocSpan$ for
brevity. On the event $(\DocSpan > \CharTime{\Csize})$, we similarly obtain
\begin{align*}
& \MeanHitsDoc \; = 
\\
& \sum_{k = 2}^{+\infty}
\left(
	\sum_{\iReq = 2}^{k} 
	\ECSS{\II{\Req_\iReq - \Req_{\iReq - 1} < \CharTime{\Csize}}}
	     {\Doc}{\NReqs_\Doc = k}
\right)
e^{-\DocReqRate \DocSpan} 
\frac{\left( \DocReqRate \DocSpan \right)^k}{k!} \; =
\\
& 
\sum_{k = 2}^{+\infty}
(k-1)
	\left [1 - \left(1 - \frac{\CharTime{\Csize}}{\DocSpan} \right)^k \right]
e^{-\DocReqRate \DocSpan} 
\frac{\left( \DocReqRate \DocSpan \right)^k}{k!} \; =
\\
& \sum_{k = 2}^{+\infty}
(k-1) \left[
	 \frac{\left( \DocReqRate \DocSpan \right)^k}{k!} 
	- \frac{(\DocReqRate (\DocSpan - \CharTime{\Csize}))^k}{k!} \right]
e^{-\DocReqRate \DocSpan} \; =
\\
& 
\DocReqRate \DocSpan - 1 + e^{-\DocReqRate \DocSpan} 
 - e^{-\DocReqRate \CharTime{\Csize}}(\DocReqRate (\DocSpan - \CharTime{\Csize}) - 1 + e^{- \DocReqRate (\DocSpan -\CharTime{\Csize}})) \; =
\\
& 
(\DocReqRate \DocSpan - 1)(1 - e^{-\DocReqRate \CharTime{\Csize}}) 
 + \DocReqRate \CharTime{\Csize} e^{-\DocReqRate \CharTime{\Csize}}
\end{align*}
which concludes the proof.
\end{IEEEproof}



\end{document}